\documentstyle[aps,prb,psfig]{revtex}

\begin{document}

\twocolumn[\hsize\textwidth\columnwidth\hsize\csname @twocolumnfalse\endcsname
\title{ Evidence for Stimulated Scattering of Excitons into Microcavity
Polaritons} \author{F. Tassone, R. Huang, Y. Yamamoto} \address{ Erato
Quantum Fluctuation Project, E. L. Ginzton Laboratory, Stanford
University, Stanford, California 94305, U.S.A. }

\maketitle

\begin{abstract}

The wavefunction of a collection of identical quantum particles of
integer spin (bosons) is even under exchange of the coordinates of any
two of them.\cite{identical,identical2}  This symmetrization rule of
quantum mechanics leads to stimulation of the scattering rate into
final states whose population is larger than one
.\cite{stimulation} For massless photons, it is routinely observed in
the laser.  For massive bosons, most previous efforts concentrated in
realizing the largely populated equilibrium Bose-Einstein condensate,
and observing stimulated emission into it.\cite{Mysrowicz}  Here we
show that we do not need to produce such a condensate to observe final
state stimulation. Instead, we take advantage of the highly
non-equilibrium exciton reservoir and of the large polariton
populations produced by external laser action on a semiconductor
microcavity, and measure final state stimulation of the scattering
rate of these excitons into the polaritons.\end{abstract}
]

The exciton is a composed of an electron and a hole (fermions), and is
the equivalent of an hydrogen atom in a semiconductor. Having integer
spin, it behaves as a boson, but due to its internal structure, it is
not ideal. This non-ideality can be modeled as an effective
interaction between fully bosonic excitons. This approximation breaks
down when $n_{exc}\simeq a_B^{-d}$ (Mott density),\cite{bosonic} where
$a_B$ is the exciton Bohr radius, d the dimensionality of the
system. Here we are considering two-dimensional excitons confined in a
GaAs 200\AA\ quantum well embedded in a microcavity which confines the
photons (inset, Fig. 1). Due to confinement, the photon is
two-dimensional and acquires a finite mass.  Typical dispersion
relations of excitons and cavity photons are shown in Fig. 1. The
exciton and photon are degenerate at the in-plane momentum $\bf
k$=0. Strong dipole interaction between them results in the formation
of mixed modes, the exciton-polaritons.\cite{micro} Thus polaritons
are also interacting bosons below the Mott density. The typical $k=0$
splitting between upper (UP) and lower polariton (LP) in our structure
is 3.6 meV. The photon and exciton mass differ by almost 4 orders of
magnitude.  Already at small $\bf k$ they become essentially
uncoupled: the UP becomes photon-like, the LP exciton-like. Hereafter,
we use UP and LP to refer to the strongly coupled modes at $\bf
k\simeq$0, and excitons to refer to the exciton-like LP at larger
$k$. Due to finite transmission of the mirrors enclosing the cavity,
polaritons and excitons of given $k$ can be excited by external laser
beams, with $k=\omega/c\sin\theta$, where $\theta$ is the external
incidence angle, $\hbar\omega$ the photon energy, c the velocity of
light.  This external world coupling also results in spontaneous decay
of polaritons and excitons into external photons. The polariton
lifetime at $\bf k$=0 was measured $\tau_{pol}\sim$2 ps. Excitons
having $k<n\omega/c$ (n the index of refraction of the substrate) have
lifetimes well over 20 ps, while those with $k>n\omega/c$ have
infinite radiative lifetime as decay into external photon is forbidden
by energy and momentum conservation.\cite{agranovich} 

The
thermalization process of these particles involves scattering with the
lattice vibrations (phonons). As a result of the largely different
masses of excitons and polaritons, the latter have a much smaller
scattering rate with phonons than the former. LP do not have time to
scatter with phonons before they decay back into external photons, and
do not thermalize to the lattice temperature. Instead, excitons do
approximately thermalize.\cite{tassone.prb}
\begin{figure} 

\centerline{
\psfig{figure=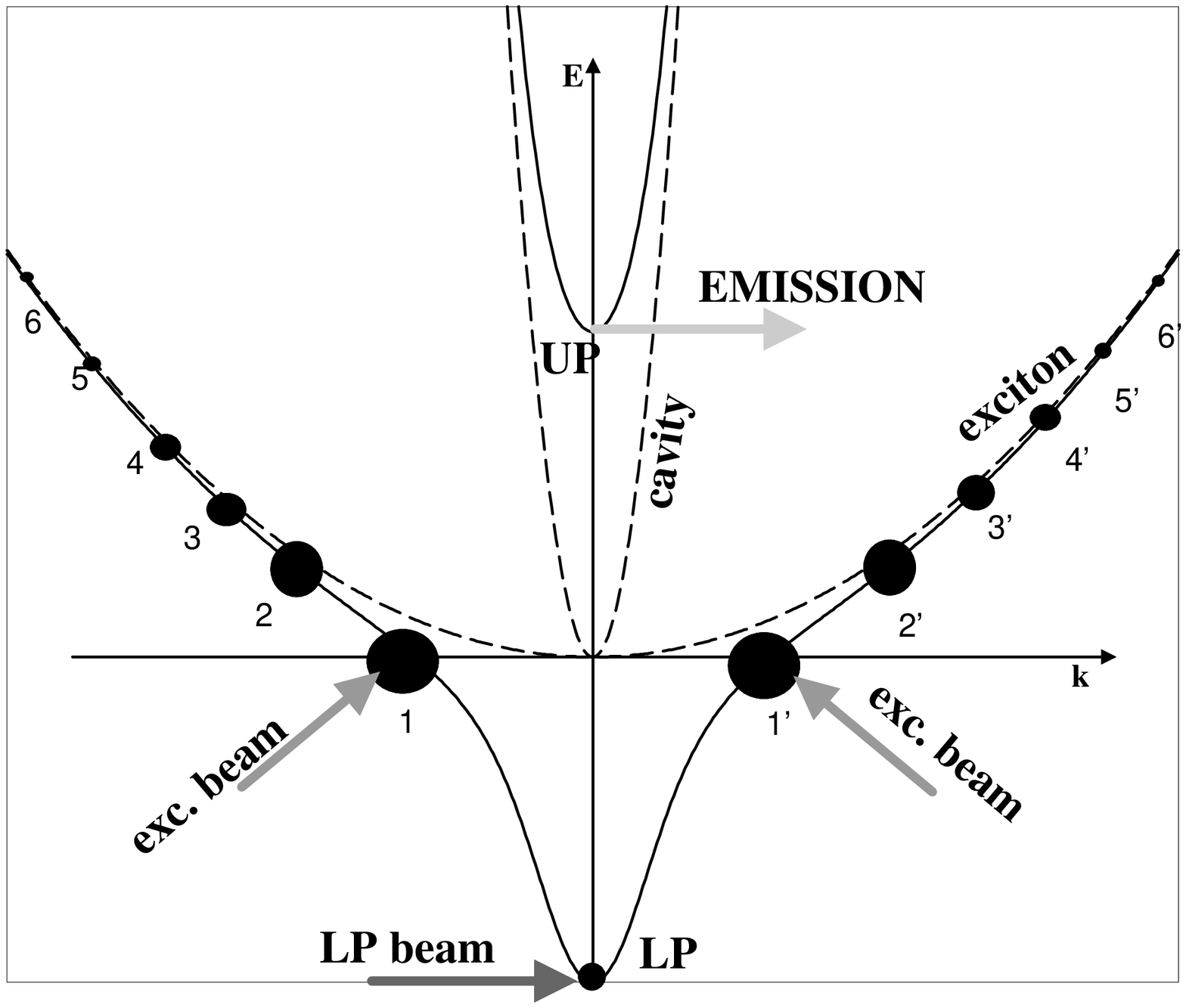,width=7.5truecm}
\hspace{-5.8truecm}\raisebox{2.75truecm}
{\psfig{figure=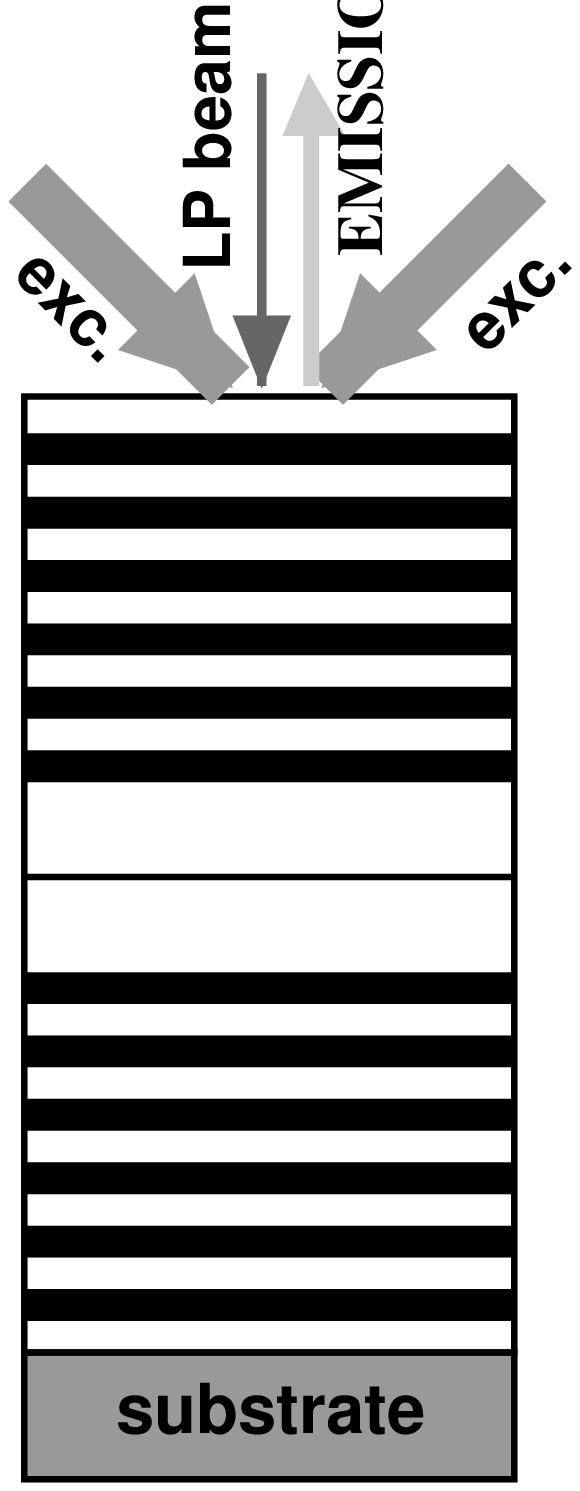,width=1.5truecm}}
\hspace{4.9cm}
}

\caption{
Dispersion of exciton-polaritons. Exciton (not to scale) and cavity
photon (dashed lines) dispersion also shown.  The exciton population
is depicted by solid circles of proportional size, also
labelled. Typical scatterings are ($c_1$) 6$\rightarrow$UP$\pm$phonon;
($c_2$) 1+6'$\rightarrow$UP+4; ($c_2^\prime$) 1+1'$\rightarrow$UP+LP.
Inset: the cross section of the microcavity structure.  Quarter
wavelength dielectric stacks (distributed Bragg reflectors) confine
the photon inside the cavity, into which a GaAs quantum well,
confining the exciton, is embedded. The microcavity is grown on a
substrate and mounted in a criostat. In the experiment, two exciton
beams at large angles and a LP beam at the normal direction are
incident upon the top facet. The emission is also collected in the
normal direction, and analyzed with a CCD spectrometer.}
\end{figure}
 However, excitons are
prevented to further cool down and transform into LP by a relaxation
bottleneck, related to the above mentionned slow down of phonon
scattering.\cite{toyozawa,tassone.prb} Due to the short LP lifetime,
the resulting LP population from this relaxation process is always
small ($<10^{-4}$ for the exciton densities considered later). Thus,
injection of excitons and/or polaritons in a microcavity produces a
state of the system which is far from thermal equilibrium: the exciton
and lower polariton populations can be varied {\em
independently}. This allows to change both the initial and final state
populations of the elastic scattering process where two excitons are
scattered into the LP and UP respectively.  In particular, the rate of
emission of UP in this process is proportional to $n_{exc}^2
(1+N_{LP})(1+N_{UP})$.  Then, final state stimulation of this
scattering can be induced by injection of LP using an external laser
beam.  The resulting $N_{UP}\ll 1$, due to the weak scattering rates
and the short UP lifetime.

We realized the set-up schematically shown in the inset of Fig. 1,
where two pump Ti-sapphire laser (exciton beams) of intensity
$I_{exc}$ at large angle (55$^o$) and at the exciton energy inject an
exciton population, and a third pump semiconductor laser (LP beam) of
intensity $I_{LP}$ at normal incidence and at the LP energy injects
LP.  These laser were continous wave, focussed to a spot size $S$ of
about $(40\mu$m$)^2$. The microcavity sample was placed in a liquid He
cryostat of nominal temperature $T_{c}=$4.8K.  Emission from the UP
was collected at normal incidence and spectrally resolved from the
lower polariton emission using of a CCD spectrometer.  The typical
emission spectrum is shown in Fig. 2, for a fixed $I_{exc}$ and
varying $I_{LP}$.  
\begin{figure} \centerline{\psfig{figure=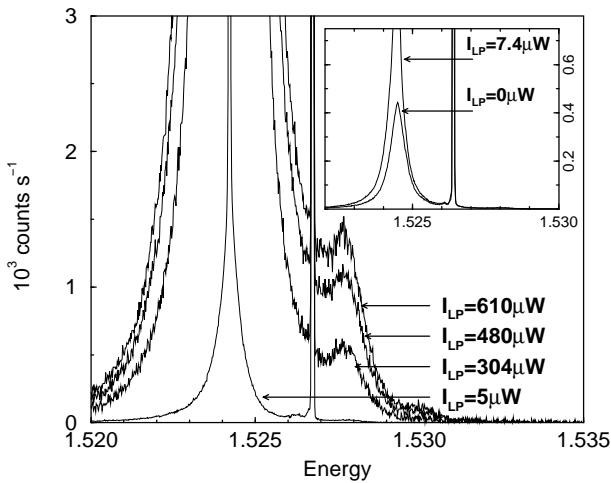,width=8truecm}}
\caption{
Raw photoluminescence emission spectrum. Here shown for different
$I_{LP}$, and fixed $I_{exc}$= 12 mW.  Stray light diffused from the
exciton beam at 55$^o$ produces a sharp peak at 1.5268 eV. Stray light
diffused from the reflected semiconductor laser (LP beam) at close
angles also shows at the LP energy, with the laser spectral shape, and
is cut-off in the scale of the figure. In the inset, the
emission for $I_{LP}=0$, and $I_{exc}$= 1 mW. A LP
population $\sim 10^{-6}$ contributed from relaxation of
excitons is deduced.}\end{figure} 
The peak of emission from the UP is well
resolved from the background emission of the LP. The LP spectrum is
dominantly the semiconductor laser spectrum.  The height of the UP
emission varies with $I_{LP}$, and features a constant full width at
half maximum close to 0.5 meV. No shifts or additional broadenings are
noticed. Moreover, this holds for the available exciton beam powers
$I_{exc}<$ 150 mW, and $I_{LP}< $60 $\mu$W. It indicates absence of
changes of polariton splitting or significant additional broadening by
scattering with other excitons.  We came to the same conclusions from
reflectivity measurement at $I_{exc}=0$ and 17mW, Fig. 3.

\begin{figure}
\centerline{\psfig{figure=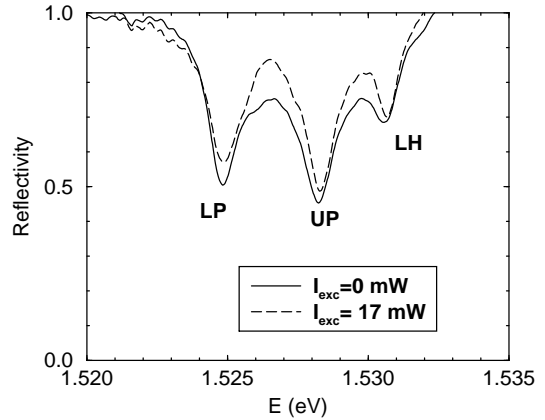,width=7truecm}}
\caption{ The
reflectivity spectra from the sample with $I_{exc}$=0, and 17 mW.
These spectra where taken using 150fs pulsed laser source centered at
1.527eV, with very weak intensity. To reduce collection time, the
spectrometer resolution was lowered to around 0.5 meV. The data was
then normalized with the gaussian shape of the source spectrum.  Due
to the very weak reflectivity signal and stronger stray light from the
17mW exciton beam, we also had to subtract spurious effects of the
tails of this signal using a Lorentzian. We then fitted the
reflectivity spectrum with Lorentzians. We did not find significant
shifts in the position of the polariton peaks, and in their
broadenings, within the spectrometer resolution. The third peak at
1.5306 eV is related to the LH exciton (4 meV above the HH
exciton). The cavity is tuned in resonance with the HH exciton, so
that the resulting LP and UP are mainly contributed from the HH
exciton and cavity photon mode, having negligible LH components. This 
component smaller than 10\%\ for both of
them. Moreover, the LH exciton is irrelevant in the low temperature
dynamics of the system, due to its higher energy, and consequent
negligible population.}
\end{figure}

The
polariton picture is thus relevant in the whole experimental range
considered. Moreover, the fact that the UP emission shows its typical
linewidth rules out any contribution from coherent processes like four
wave mixing, in which case the emission would show a much narrower
linewidth related to that of the exciton and LP lasers. This
observation strongly corroborates the dynamical picture presented
above, that substantial scattering of the excitons with the phonons
takes place.  In Fig. 4 (a) we plot typical results of the UP
emission rate as a function of $I_{exc}$, for two different $I_{LP}$.
In order to avoid any concerns about large exciton densities, or
possible local heating of the substrate, we collected an homogeneous
data set for five different $I_{LP}$ and limiting $I_{exc}<$ 30 mW. We
do not include the two curves in Fig. 4 (a) in this data set, as
different spot sizes and spectrometer resolution were used to
increase collection statistics.  We fitted the curves in the data set
with simple parabolas $dN_{UP}/dt=c_1(I_{LP})I_{exc}+c_{2}(I_{LP})
I_{exc}^2$, for each of the five different $I_{LP}$. The results for
$c_1$ and $c_2$ are shown in Fig. 4 (b). $c_1$ is independent of
$I_{LP}$, whereas $c_{2}$ is linearly dependent on it. Since the
exciton density and the LP population are proportional to $I_{exc}$
and $I_{LP}$ respectively, the exciton-exciton scattering into LP and
UP is quadratic in $I_{exc}$ and the stimulated part, linear in
$I_{LP}$.  Thus, Fig. 4 (b) is a direct experimental proof of final
state stimulation of the exciton-exciton scattering process due to
$N_{LP}>1$.

\begin{figure} 
\centerline{\psfig{figure=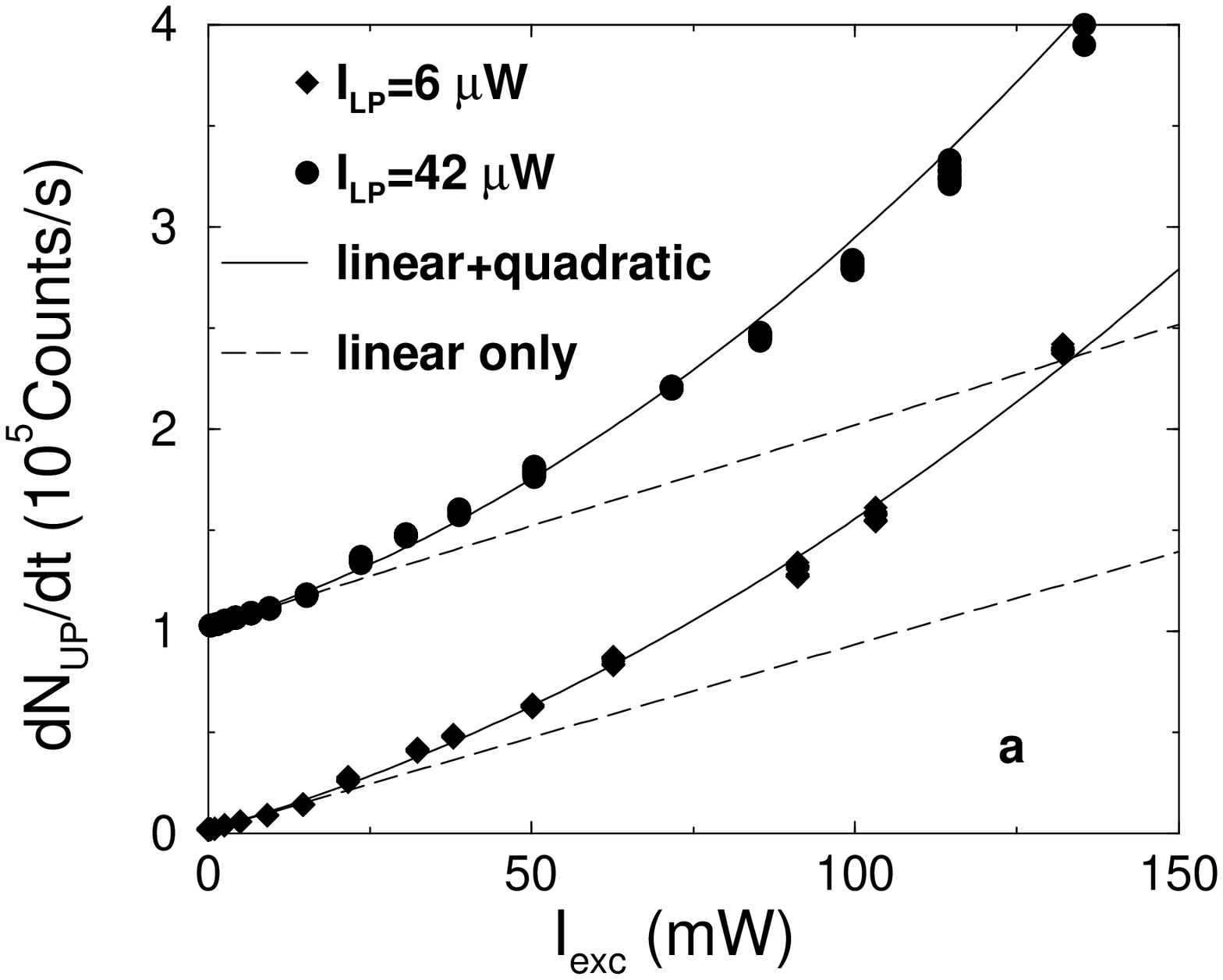,width=7truecm}}

\centerline{\psfig{figure=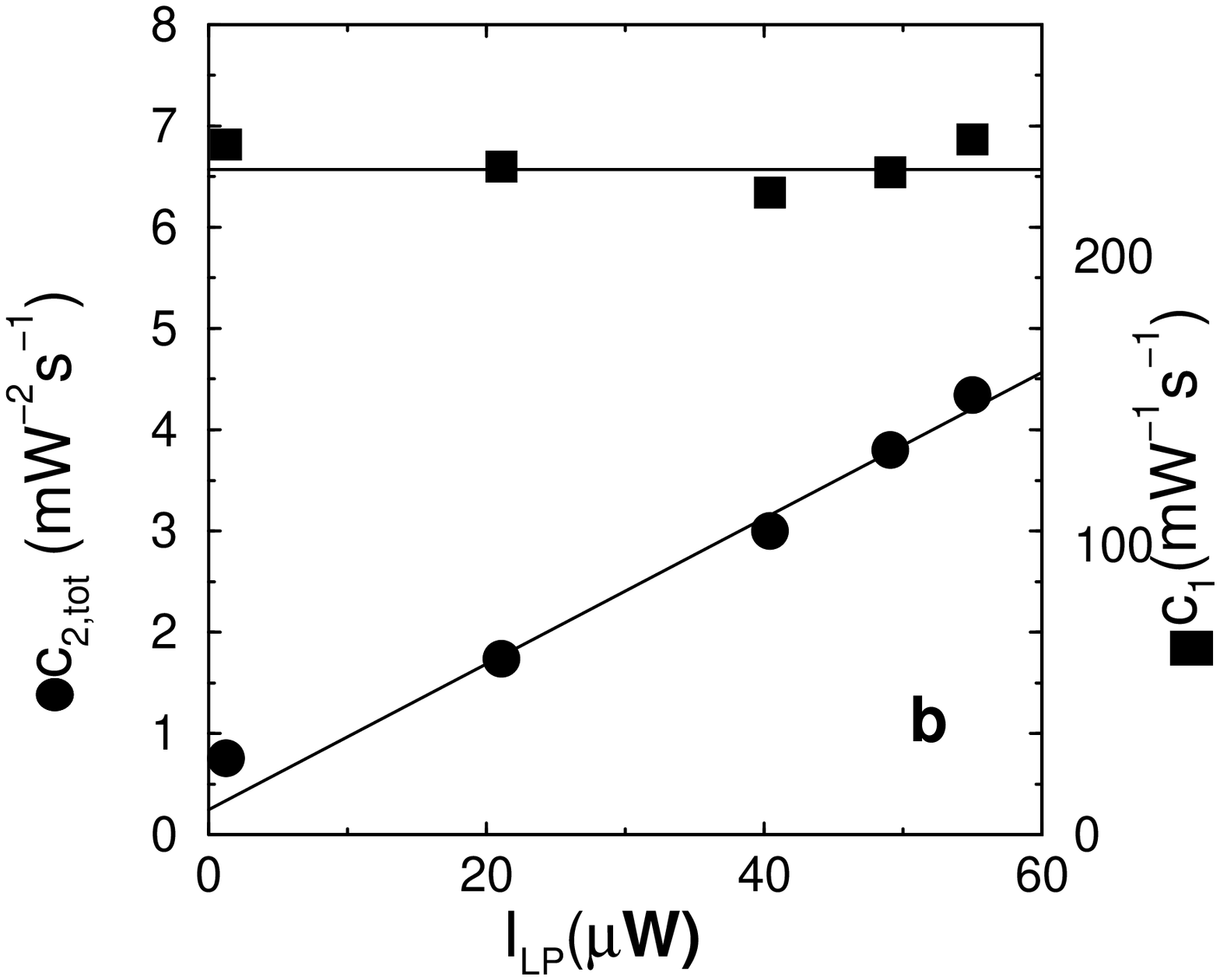,width=7truecm}}

\caption{
Upper polariton emission rate dependence on $I_{exc}$ and
$I_{LP}$.  a: as a function of $I_{exc}$, for two different $I_{LP}$,
the highest set shifted by 10$^5$ for clarity.  Dashed lines: the
linear components only. b: the linear and total quadratic dependence
on $I_{exc}$, as a function of $I_{LP}$. Solid lines: fit with a
constant for $c_1$ and with a line for $c_{2,tot}$.
}\end{figure}

Beside the exciton-exciton scattering to LP and UP, other types of
scattering take place.  At finite temperature, excitons scatter with
acoustic phonons to the UP.\cite{tassone.prb} The rate of this process
is linear in the exciton density, and independent of $I_{LP}$, thus is
at the origin of the $c_1I_{exc}$ term.  Moreover, two excitons from
the non-equilibrium reservoir may also elastically scatter into
another exciton of lower energy and an UP. This scattering is not
stimulated by LP population, but is quadratic in exciton density. We
may instead neglect UP emission from bound biexcitons produced by
absorption of exciton and LP beams, as the resulting biexciton density
is negligible.  We calculated the scattering rates, within the Fermi
golden rule, for the exciton-phonon scattering $\tilde c_1=1.1\cdot
10^{-2}$cm$^2$s$^{-1}$ using deformation potential interaction of
excitons with acoustic phonons, for the exciton+exciton to exciton+UP
$\tilde c_2=4.2\cdot 10^{-11}$cm$^4$s$^{-1}$, and for the
exciton-exciton to LP+UP $\tilde c_2^\prime= 3.9\cdot
10^{-13}$cm$^4$s$^{-1}$ using the exciton-exciton exchange
interaction.
\cite{bosonic}  In order to compare theory and experiment, we need to
establish how exciton density and LP populations are related to $
I_{exc}$ and $I_{LP}$ respectively. As the exciton-phonon scattering
is a well understood dynamical process, its theoretical prediction is
accurate. The experimental exciton-phonon scattering rate is given by
$c_1/Q I_{exc}$, where $Q\sim$1\%\ is the measured overall detection
efficiency. Apart from inaccuracies in $Q$, $c_1 I_{exc}$ is
accurately measured, $c_1=230\pm 3$mW$^{-1}$s$^{-1}$.  Thus, we have a
reasonable estimate of $n_{exc}=\beta I_{exc}$ by equating the
experimental and theoretical rates.  We obtain $\beta=Q^{-1}
c_1/\tilde c_1= 2.1\cdot 10^6$ mW$^{-1}$cm$^{-2}$. This $\beta$ is
only a factor 4 smaller than the one directly estimated from the
exciton lifetime $\tau_{exc}=100 ps$,\cite{andreani} the spot surface
$S$, and the total exciton absorption of 4$\cdot 10^{-4}$ (this small
value stemming from the high top mirror reflectivity at 55$^o$,
measured 99 \%). The agreement between the two independent estimates
is reasonable in view of significant uncertitudes in the parameters
above, and in $Q$.  We remark that $n_{exc}=$ 6$\cdot 10^7$cm$^{-2}$
at $I_{exc}=30$mW, much smaller than the Mott density for GaAs QW of
2$\cdot$10$^{11}$cm$^{-2}$.\cite{SR} At these $n_{exc}$, scattering
rates are negligible in comparison with the polariton splitting and
broadenings, consistently with the experimental findings above.  In
different experiments using similar excitation powers,\cite{boser1}
collapse of the polariton splitting was observed.\cite{boser2}
However, there free electron-hole carriers were injected at normal
incidence above the top mirror stop band (low mirror
reflectivity). Hence the resulting {\em free carrier} density was at
least 3 orders of magnitude larger than our {\em exciton} density,
considering the longer free carrier radiative lifetime. We now
estimate $N_{LP}=\gamma I_{LP}=[1 \mu{\rm W}]
\tau_{pol}=8.5\mu W^{-1}$ directly from the LP lifetime $\tau_{pol}$, 
also assuming all the incident flux converted into LP.  Using
$\beta,\gamma$ and Q, we would expect from theory to measure $c_2$=1.8
mW$^{-2}$s$^{-1}$, and $c_2^\prime$=0.14
mW$^{-2}\mu$W$^{-1}$s$^{-1}$. Instead, we measured $c_{2}=.25\pm .17$
mW$^{-2}$s$^{-1}$, $c^\prime_{2}=.072\pm
.005$mW$^{-2}\mu$W$^{-1}$s$^{-1}$. The agreement is only qualitative,
yet shows we are grasping the fundamental physics of the system.

A relevant aspect of the real system is the presence of QW interface
roughness and alloy disorder. It may lead to a decrease of the
exciton-exciton scattering efficiency, as elastic exciton scattering
on roughness perturbs the free exciton trajectories. The experimental
results here indirectly show that this diffusion is instead not much
relevant for the LP, as large populations in the same quantum state
are required to observe stimulated scattering. This result is both due
to the short cavity photon lifetime and to its small mass. As the LP
is a 1:1 mixture of exciton and cavity photon, and both the cavity and
LP linewidths are the same, 0.5 meV, we conclude that at least 50 \%\
of the injected LP remain in the same quantum state, and do not
diffuse to other k-states before decaying.  Roughness is also at the
origin of a certain surface density of strongly localized, and low
energy, excitons.  Due to strong confinement, they loose bosonic
character, featuring a typical fermionic saturation behaviour in their
population dynamics. Through this saturation dynamics, the LP beam may
modulate the exciton density. However, we can exclude it, as the
linear phonon scattering, proportional to $n_{exc}$, does not vary
with $I_{LP}$, Fig. 4 (b), $c_1$ coefficient. In conclusion, our
observations of stimulated emission prove that polariton beams
are practical realizations of massive coherent waves, and opens the
possibility of studying their peculiar bosonic properties in
conventional semiconductor structures.

 \end{document}